\documentclass{ws-ijmpa}
\usepackage[super,compress]{cite}
\usepackage{graphicx}
\begin{document}
\markboth{A. Andrianov, Y. Elmahalawy  \& A. Starodubtsev}{Cylindrically symmetric 2+1 gravity in terms of global variables}

%
\catchline{}{}{}{}{}
%

\title{Cylindrically symmetric 2+1 gravity in terms of global variables: quantum dynamics}

\author{\footnotesize Alexander A. Andrianov}

\address{St.Petersburg state university, Petergof ul. Ulyanovskaya, 1
\\ St.Petersburg, 198504, Russia \\
sashaandrianov@gmail.com}

\author{\footnotesize Yasser Elmahalawy}

\address{St.Petersburg state university, Petergof ul. Ulyanovskaya, 1
\\ St.Petersburg, 198504, Russia \\
yasserreda99@gmail.com}

\author{\footnotesize Artem Starodubtsev}

\address{St.Petersburg state university, Petergof ul. Ulyanovskaya, 1
\\ St.Petersburg, 198504, Russia \\
artremstarodubtsev@gmail.com}

\maketitle

\begin{history}
\received{Day Month Year}
\revised{Day Month Year}
\end{history}

\begin{abstract}
We perform quantization of a model in which gravity is coupled to a circular dust shell in 2+1 spacetime dimensions.
Canonical analysis shows that momentum space of this model is $ADS^2$-space, and the global chart for it is provided by the Euler angles.
In quantum kinematics, this results in non-commutativity in coordinate space and discreteness of the shell radius in timelike region, which includes the collapse point.
At the level of quantum dynamics, we find transition amplitudes between zero and non-zero eigenvalues of the shell radius, which describe the rate of gravitational collapse (bounce). Their values are everywhere finite, which could be interpreted as resolution of the central singularity.

\end{abstract}

\ccode{PACS numbers: 04.60.-m, 04.60.Kz, 04.60.Ds}


\section{Introduction}

It has long been believed that quantum theory of gravity could regularize its singularities. Perhaps the simplest model for investigating this possibility is gravity coupled to a spherically symmetric dust shell.
There is a variety of works studying such models both on classical \cite{israel1,kuchar,hajicek } and quantum \cite{louko,vb,hk,vb1} level.
In some of the versions of quantum theory the central singularity is removed.\cite{vb,hk,vb1} However, the above results do not always agree with each other. Apart from quantization ambiguity, the other possible reason for that is a complicated phase space structure of the model. Different quantum theories could arise on different sectors of such phase space.

In such situation the common wisdom is that the wavefunction of a quantum theory has to be defined on all possible configurations, independently of whether they are classically accessible or not. In our previous paper Ref. \citen{aes}, we found canonical variables which provides the global chart for the entire phase space of such a model in 2+1 spacetime dimensions. At the kinematical level, quantization in terms of these variables leads to some non-trivial consequences, such as spacetime discreteness. This paper mostly concerns the dynamics of this model.

In section 2 we review the kinematical results obtained previously and supply some explicit results which will be used in  studying dynamics.

In section 3 we find the expression for the evolution operator of the model and calculate its matrix elements which describe the transition amplitudes between different shell radii, including zero radius. These amplitudes turn out to be everywhere finite, which means that the model is singularity-free.

\section{Canonical Analysis and Quantum Kinematics}

We use the Einstein-Cartan formulation of gravity, which is similar to gauge theory, and the matter (dust shell) is introduced as an extrinsic charge for the gauge field.
The basic variables are the triad $e_\mu=e_\mu^a\gamma_a$ and the connection $\omega_\mu^{ab}\gamma_a\gamma_b$, where $\gamma^a$ are generators of sl(2)-algebra.
The action reads:
\begin{equation}
\label{action3d}
S=\int_M d^3 x \epsilon^{\mu \nu \rho} Tr (e_\mu R_{\nu \rho}) +  \int_{l} Tr(K e_\mu) dx^\mu d\phi,
\end{equation}
where $R_{\nu \rho}$ is the curvature of $\omega_\rho$,  $l$ is the shell worldsheet and $K=M \gamma_0$ -- a fixed element of sl(2)-algebra, $M$ is the bare mass of the shell.

Following Ref.\citen{aes}, we discretize the shell  (describe it as an ensemble of $N$ point particles), and for each point particle apply the results of Refs. \citen{thooft,mw}.

After performing hamiltonian reduction and taking limit $N \rightarrow \infty$ we obtain the following kinetic term in the action
\begin{equation}
\label{kinetic}
S_{kinetic}=
 \int_{R } \xi^a Tr( \gamma_a U^{-1}\dot U ),
\end{equation}
where $\xi^a$ is the shell coordinates (having only temporal and radial components), and $U$ is an SL(2) group valued holonomy of connection $\omega$ around the shell, which is the product of holonomies around every particle. Holonomy $U$ provides the global parametrization of the momentum space of the shell, which has $ADS^2$-geometry.

Quantization is performed in momentum representation, where the wavefunction $\Psi$ is a function of holonomy of the shell parametrized by the Euler angles $U=\exp(\frac{\rho}{2}\gamma_0)\exp(\chi \gamma_1)\exp(\frac{\rho}{2}\gamma_0)$. It is periodic in $\rho$, $\Psi(\rho,\chi)=\Psi(\rho+2\pi,\chi)$.

We calculate the spectrum of the follwing two commuting variables: time coordinate $t=\xi^0$ and the shell radius $R^2=\xi_a\xi^a$.
In quantum theory they are represented by derivative w.r.t. $\rho$, $\hat t = i\frac{\partial}{\partial \rho}$, and the Beltrami-Laplace operator on $ADS_2$, $\hat R^2 =-\Delta_{ADS^2}$. The spectrum of time operator is discrete due to periodicity in $\rho$, $\hat t |t, R^2 \rangle=t |t, R^2 \rangle$, $t\in\mathbb{Z}$. The spectrum of the shell radius in spacelike region, $R^2>0$, is continuous and separated from zero, $\hat R^2 |t, \lambda \rangle=(\lambda^2+1/4)|t, \lambda \rangle$, $\lambda \in\mathbb{R}$.

The corresponding eigenfunctions are $\langle\rho, \chi | t,\lambda \rangle=e^{it\rho}(P^t_{i\lambda}(\tanh\chi)+Q^t_{i\lambda}(\tanh\chi))/\sqrt{\cosh \chi}$, where $P^t_{i\lambda}$ and $Q^t_{i\lambda}$ are first and second kind generalized Legendre functions respectively.
In timelike region, $R^2\leq 0$, the spectrum of the shell radius is discrete $\hat R^2 |t, l \rangle=-l(l+1)|t, l \rangle$, $l\in\mathbb{Z}^+$. The corresponding eigenfunctions are $\langle\rho, \chi | t,l \rangle=e^{it\rho}P^t_{l}(\tanh \chi)/\sqrt{\cosh \chi}$, where $P^t_{l}$  are  associated Legendre polynomials. As we see, the central singularity point, $R^2=0$, belongs to the discrete spectrum.

\section{Quantum Dynamics}	

To describe the dynamics of the model one has to take into account the hamiltonian constraint which relate momenta of the shell to its bare mass.\cite{aes} The group-valued momentum variable $U$ can be expressed in terms of the bare mass $M$ and radial boost parameter $\bar\chi$ by multiplying holonomies of all the particles composing the shell $u_i$
\begin{equation}\label{ufin}
U=\prod u_i =\exp(2\pi ((1-M \cosh \bar\chi) \gamma_0 + M \sinh \bar\chi \gamma_2) ).
\end{equation}

This matrix relation contains two independent equations and a free parameter, $\bar\chi$. By excluding this parameter one can obtain a single equation which is the Hamiltonian constraint. Because of complications the form of the constraint is derived approximately by interpolation between $\bar\chi\rightarrow 0$ and $\bar\chi\rightarrow \infty$ limits
\begin{align}
\label{discrete2}
Tr(U) & =\cosh \chi\cos\rho \nonumber \\  & \approx  \cos\bigg(2\pi  \sqrt{1+M^2-M\sqrt{1+\frac{(1-M)^2}{M^2\sin^2(2\pi(1-M))}\chi^2}}\bigg).
\end{align}

This expression is valid only for finite $M$. Its quantum version is an analog of Klein-Gordon equation for relativistic particle.
In a skew representation (coordinate in time variable and momentum in radial variable) it has the form:
\begin{align}
\label{discrete3}
\Psi(t+1,\chi) +\Psi(t-1,\chi) & = H(\chi)\Psi(t,\chi),
\end{align}
where
\begin{align}
\label{discrete4}
H(\chi) & = \frac{ \cos\bigg(2\pi  \sqrt{1+M^2-M\sqrt{1+\frac{(1-M)^2}{M^2\sin^2(2\pi(1-M))}\chi^2}}\bigg) }{\cosh(\chi)}.
\end{align}

Due to periodicity in energy variable this is not a differential equation, but a finite difference equation, which is in agreement with discreteness of time found in the previous section. By separation positive  and negative frequency parts this equation can be put in Shroedinger-like form:
\begin{align}
\Psi(t\pm 1,\chi)\equiv \hat U^{\pm 1} \Psi(t,\chi)  =  \Big( H(\chi) \pm \sqrt{2 H^{2}(\chi)-2}\Big)\Psi(t,\chi),
\end{align}
where $\hat U$ is the evolution operator for one step in time.

\begin{figure}[h!]
\centerline{\includegraphics[width=3.5in]{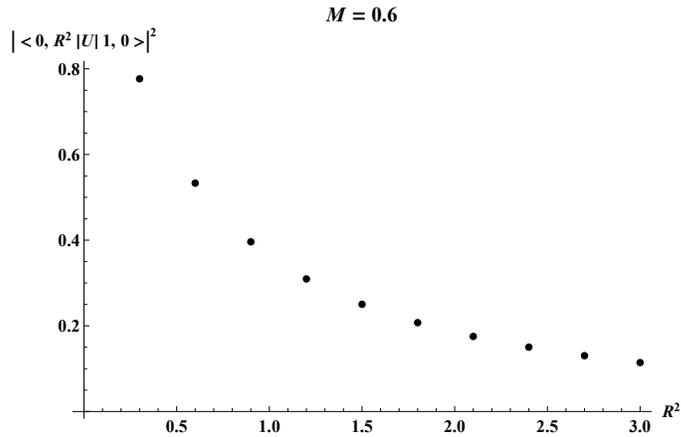}}
\vspace*{8pt}
\caption{Absolute value of matrix element $\langle 1,0 |\hat U|0,R^2\rangle$ describing gravitational collapse rate. \protect\label{fig1}}
\end{figure}

The gravitational collapse rate is described by a matrix element of the evolution operator $\hat U$ between eigenstates of the shell radius $R$ for positive (spacelike) $R^2$ and for zero $R^2$, $\langle t+1,0 |\hat U|t,R^2\rangle$.  On the plot Fig. \ref{fig1}, we show the result of numerical calculation of the absolute value of matrix element $\langle 1,0 |\hat U|0,R^2\rangle$ for different $R^2$. One can see that this value is everywhere finite, including minimal possible positive eigenvalue of the shell radius, $R^2=1/4$, which can be interpreted as resolution of the central singularity.

\section*{Acknowledgments}
We would like to thank the organizers of 10th Friedmann seminar. AA is grateful to the organizers of V Russian-Ibertian Congress for support. This work was supported in parts by RBFR grant 18-02-00264 (A.A. and A.S.). AA is also supported by the SPbSU travel grant 41327333.


\end{document}